\documentclass[a4paper,aps,prd,onecolumn,preprintnumbers,showpacs,superscriptaddress,nofootinbib]{revtex4}

\usepackage{graphics,graphicx,epsfig}
\usepackage{amsfonts,amsmath,amssymb}

\begin{document}

\title{Null geodesics and shadow of a rotating black hole\\ in extended Chern-Simons modified gravity}
\author{Leonardo Amarilla}
\email{yellow@df.uba.ar} \affiliation{Facultad de Ciencias
Astron\'omicas y Geof\'{\i}sicas, Universidad Nacional de La Plata,
Paseo del Bosque, 1900, La Plata, Argentina.}
\affiliation{Departamento de F\'{\i}sica, Facultad de Ciencias Exactas y Naturales, Universidad de Buenos Aires, Ciudad Universitaria, Pabell\'on 1, 1428, Buenos Aires, Argentina.}
\author{Ernesto F. Eiroa} \email{eiroa@iafe.uba.ar} \affiliation{Instituto de Astronom\'{\i}a y F\'{\i}sica del Espacio, C.C. 67 Suc. 28, 1428, Buenos Aires, Argentina.} \affiliation{Departamento de F\'{\i}sica, Facultad de Ciencias Exactas y Naturales, Universidad de Buenos Aires, Ciudad Universitaria, Pabell\'on 1, 1428, Buenos Aires, Argentina.}
\author{Gaston Giribet}
\email{gaston@df.uba.ar}
\affiliation{Departamento de F\'{\i}sica, Facultad de Ciencias Exactas y Naturales, Universidad de Buenos Aires, Ciudad Universitaria, Pabell\'on 1, 1428, Buenos Aires, Argentina.}


\begin{abstract}
The Chern-Simons modification to general relativity in four dimensions consists of adding to the Einstein-Hilbert term a scalar field that couples to the first class Pontryagin density. In this theory, which has attracted considerable attention recently, the Schwarzschild metric persists as an exact solution, and this is why this model resists several observational constraints. In contrast, the spinning black hole solution of the theory is not given by the Kerr metric but by a modification of it, so far only known for slow rotation and small coupling constant. In the present paper, we show that, in this approximation, the null geodesic equation can be integrated, and this allows us to investigate the shadow cast by a black hole. We discuss how, in addition to the angular momentum of the solution,
the coupling to the Chern-Simons term deforms the shape of the shadow.
\end{abstract}

\maketitle

\section{Introduction}

The Chern-Simons (CS) modification to Einstein general relativity (GR) consists of augmenting Einstein-Hilbert action by adding a parity violating gravitational term which couples a scalar field $\varphi $ to the first-class Pontryagin density $\ast R_{\mu \nu \rho \sigma }R^{\mu \nu \rho \sigma }$, where $\ast R_{\alpha \beta \rho \sigma }=\frac{1}{2}\varepsilon_{\rho \sigma }^{~~~ \mu \nu}R_{\alpha \beta \mu \nu }$ is the dual Riemann tensor \cite{JP}. With this modification, the gravitational action in the absence of matter takes the form
\begin{equation}
S=\kappa \int dx^{4}\sqrt{-g}R-\frac{1 }{2}\int dx^{4}\sqrt{-g}\left( \nabla \varphi \right) ^{2}+\frac{\gamma }{4}\int dx^{4}\sqrt{-g} \varphi \, \ast R_{\mu \nu \rho \sigma }R^{\nu \mu \rho \sigma },  \label{action}
\end{equation}
where $\gamma$ is the coupling constant and $\kappa =(16\pi)^{-1}$ (throughout the article we adopt units such that $G=c=1$). The quadratic term in the action (\ref{action}) can be thought of as a gravitational parity violating analogue of the axion term. It is usual to motivate such a term from string theory as a similar term appears in the string low energy effective action and is related to the string anomaly cancellation. However, in the string theory context the natural order of magnitude of the predicted coupling constant is several order of magnitudes lower than the one required for the effects to be observed in astrophysics \cite{review}. This CS modification to Einstein theory has attracted considerable attention recently. In particular, this theory has been considered in a phenomenological context in cosmology and relativistic astrophysics; see \cite{review} and references therein. Observational constraints on the CS coupling $\gamma $ coming from spinning compact objects were given in \cite{YP}, where the bound ${\gamma^2}/{\kappa}< 5\times 10^{28} \, \mathrm{m}^4$ was obtained. It was argued in \cite{soyu} that gravitational wave detection could eventually improve this bound on $\gamma $ in a couple of orders of magnitude. More recently, in \cite{yupsal} it was discussed how the observation of slowly rotating neutron stars would yield bounds on $\gamma $ that are three order of magnitudes stronger than the one mentioned above, which comes from pulsar observations. In \cite{molina}, the gravitational perturbations of a Schwarzschild black hole in dynamical CS gravity were studied, and it was shown that the gravitational oscillation modes carry the imprints of the coupling to the scalar field, then the theory could be tested with gravitational wave detectors.

It is known \cite{Grumiller,konno2,YP,konno} that the introduction of a CS modification in the gravitational action yields a modification of the spinning black hole geometry. This is due to the fact that, unlike Schwarzschild solution, Kerr solution has a non-vanishing Pontryagin density. The spinning black hole solution to the theory (\ref{action}) is known only in the slowly rotating approximation and for small coupling constant $\gamma $ \cite{YP,konno}, and here we investigate its strong field regime by analyzing the trajectories of photons in its vicinity.

As black holes are essentially non-emitting objects, it is of interest the study of the null geodesics around them, in which photons coming form other sources move, to obtain information about these objects. In particular, gravitational lensing by black holes has received considerable attention in the last few years, mainly because of the fact of the strong evidence about the presence of supermassive black holes at the center of galaxies. A useful tool to study black hole gravitational lenses is the strong deflection limit, which is an approximate analytical method for obtaining the positions, magnifications, and time delays of the images. It was introduced by Darwin \cite{darwin} for the Schwarzschild geometry, rediscovered several times \cite{otros}, extended to Reissner-Nordstr\"om geometry \cite{eiroto}, and to any spherically symmetric black holes \cite{boz}. Numerical studies \cite{numerical} were performed as well. Kerr black holes were also analyzed in the strong deflection limit \cite{bozza1,bozza2,vazquez}.  Another related aspect that was considered, with the intention of measuring the properties of astrophysical black holes, are the shadows cast by rotating ones \cite{bardeen,chandra}, which present an optical deformation caused by the spin, instead of being circles as in the non-rotating case. This subject have been recently re-examined by several authors \cite{devries,takahashi,bozza2,hioki,bambi,maeda} due to the expectation that the direct observation of black holes will be possible in the near future. More details about these topics, additional references and a discussion of the observational perspectives can be found in the review article \cite{bozzareview}. 

In this paper, we study how the introduction of a CS term in the gravitational action modifies the null geodesics structure and the shadow produced by a spinning black hole. The article is organized as follows: In Section 2 we review the slowly spinning limit of a rotating black hole solution in CS modified gravity, for a small coupling constant; in Section 3, we study the null geodesics around the black hole and integrate the generic photon orbits. In Section 4, we find the shape of the shadow and, finally, in Section 5, we discuss the results obtained.

\section{Black holes in Chern-Simons modified gravity}

As pointed out above, we will study how the null geodesics and the shadow produced by a rotating black hole get modified by the introduction of a CS term in the gravitational action. First, let us analyze the spinning solution of the theory (\ref{action}).

The equations of motion coming from the action (\ref{action}) take the form
\begin{equation}
\nabla _{\mu }\nabla ^{\mu }\varphi +\frac{\gamma }{4}\, \ast R_{\mu \nu \rho \sigma }R^{\nu \mu \rho \sigma }=0  \label{A}
\end{equation}
and
\begin{equation}
R_{\mu \nu }-\frac{1}{2}Rg_{\mu \nu }+\frac{\gamma }{\kappa } C_{\mu \nu }=
\frac{1}{2\kappa }\nabla _{\mu }\varphi \nabla _{\nu }\varphi -\frac{1}{4\kappa }g_{\mu
\nu }\nabla _{\rho }\varphi \nabla ^{\rho }\varphi ,  \label{B}
\end{equation}
where the traceless tensor $C_{\mu \nu }$ is given by
\begin{equation*}
C^{\mu \nu }=\nabla _{\delta }\varphi \ \varepsilon ^{\delta \rho \sigma \mu }\nabla _{\sigma }R_{\rho }^{\nu }+\nabla _{\delta }\nabla _{\rho }\varphi \, \ast R^{\rho \mu \nu \delta }+(\mu \leftrightarrow \nu ).
\end{equation*}
The Pontryagin density can be written as the exterior derivative of a CS
form, namely
\begin{equation*}
\ast R_{\mu \nu \rho \sigma }R^{\mu \nu \rho \sigma }=2\nabla _{\mu }\
\varepsilon ^{\mu \alpha \beta \lambda }\left( \Gamma _{\alpha \rho
}^{\delta }\partial _{\beta }\Gamma _{\lambda \delta }^{\rho }+\frac{2}{3}
\Gamma _{\alpha \rho }^{\delta }\Gamma _{\beta \sigma }^{\rho }\Gamma
_{\lambda \delta }^{\sigma }\right) ,
\end{equation*}
and this yields the conservation of a topological current. This relation to
the three-dimensional gravitational CS term is precisely the reason why the
theory (\ref{action}) receives the name of Chern-Simons theory, even if it
sounds curious in the context of four dimensions. This makes the theory defined by (\ref{action}) specially related to the three-dimensional topologically massive gravity \cite{TMG}.

Now, let us move to discuss the spinning solution to the field equations (\ref{A}) and (\ref{B}). This solution is only known for slow rotation and small coupling constant approximation, and it was recently found in references \cite{YP,konno}. This corresponds to a perturbation of the Kerr solution of GR. So, let us recall the Kerr solution. Written in Boyer-Lindsquit coordinates, Kerr metric reads
\begin{equation*}
ds^{2}_{\text{K}}=-\left( 1-\frac{2M}{\rho ^{2} }r\right) dt^{2}+\frac{\rho
^{2}}{\Delta } dr^{2}+\rho ^{2}d\theta ^{2}-\frac{4Mra\sin ^{2}\theta }{\rho
^{2}}dtd\phi +\frac{A\sin ^{2}\theta }{\rho ^{2}}d\phi ^{2},
\end{equation*}
with
\begin{equation*}
\Delta =r^{2}-2Mr+a^{2},\quad \rho ^{2}=r^{2}+a^{2}\cos ^{2}\theta ,\quad
A=(r^{2}+a^{2})^{2}-\Delta a^{2}\sin ^{2}\theta .
\end{equation*}
The parameter $M$ is the mass of the object, while the spin parameter $a=J/M$ is given in terms of its angular momentum $J$. In the slowly spinning approximation $a\ll M$, the Kerr metric takes the form
\begin{eqnarray*}
ds^{2}_{\text{SK}} &=& -\left( B+\frac{2a^{2}M}{r^{3}}\cos ^{2}\theta \right) dt^{2}+ \frac{1}{B^{2}}\left[ B-\frac{a^{2}}{r^{2}}(1-B\cos ^{2}\theta ) \right] dr^{2}+(r^{2}+a^{2}\cos ^{2}\theta )d\theta ^{2}\\
&& -\frac{4M}{r} a\sin ^{2}\theta dtd\theta + \left[ r^2+a^2\left( 1+\frac{2M}{r}\sin ^2 \theta \right) \right]\sin ^2 \theta d\phi ^{2},
\end{eqnarray*}
where $B=1-2M/r$. On the other hand, if a CS term is added to the Einstein-Hilbert action (that is, considering $\gamma \neq 0$ in Eqs. (\ref{A}) and (\ref{B})), then the spinning solution of GR gets modified. The metric corresponding to the solution of the modified theory, in the slowly spinning and small $\gamma$ approximation, has the form \cite{YP}
\begin{eqnarray}
ds^{2} =ds^{2}_{\text{SK}} + \frac{5\gamma ^{2}}{4\kappa r^{4}}\left(1+\frac{12M}{7r} + \frac{27M^{2}}{10r^{2}} \right) a\sin ^{2}\theta dtd\theta ,
\label{metrica}
\end{eqnarray}
while the configuration corresponding to the scalar field $\varphi $ is
given by
\begin{equation}
\varphi =\left( \frac{5}{2}+\frac{5M}{r}+\frac{9M^{2}}{r^{2}}\right) \frac{\gamma a\cos \theta }{4Mr^{2}}.  
\label{escalar}
\end{equation}
From Eq. (\ref{metrica}) we observe that the off-diagonal component of the
metric receives contributions of order $\mathcal{O} (a\gamma ^{2})$. This
produces a weakened dragging effect and, as we will show below, this also
alters the null geodesics structure stamping its imprint on the shadows of
spinning compact objects.

\section{Null geodesics and photon orbits}

Lets us analyze the null geodesics around the black hole. For simplicity,
from now on we adimensionalize all quantities with the mass of the black
hole, i.e. we replace $r/M$ by $r$, $a/M$ by $a$, $\gamma /M^2$ by $\gamma$, etc. (which is equivalent to put $M=1$ in all equations).

\subsection{Equatorial photon orbits}

The equation for the trajectories of photons in the equatorial plane ($\theta=\pi /2$) can be obtained from the condition $u_{\mu}u^{\mu}=0$, where the contraction of the four-velocity $u^{\mu}$ is calculated using the metric (\ref{metrica}). In this case, we have
\begin{equation}  \label{dradtau}
\frac{1}{L^{2}}\left(\frac{dr}{d\lambda}\right)^{2}=\frac{1}{b^{2}}-W_{\text{eff}}(r),
\end{equation}
where $\lambda$ is the affine parameter, $-u_{0}=E$ is the energy, $u_{\phi}=L$ is the angular momentum in the direction of the axis of symmetry of the black hole, and
$b=L/E$ is the impact parameter. For slow rotation, the parameter $a$ is small. In this case, we approximate the effective potential $W_{\text{eff}}$ by its Taylor expansion to order $a\gamma^2$:
\begin{equation}  \label{Weff}
W_{\text{eff}}(u,b,l)=u^{2}-2u^{3}+4\frac{a}{b}u^{3}-\frac{a^{2}}{b^{2}}
u^{2}(1+2u)-\frac{2\pi u^{6}(70+120u+189u^{2})}{7 b}a\gamma^{2},
\end{equation}
where $u=1/r$. The limit $\gamma \rightarrow 0$ gives
\begin{equation}  \label{limalphacero}
\lim_{\gamma \rightarrow 0} W_{\text{eff}}=u^{2}-2u^{3}+4\frac{a}{b}u^{3}-
\frac{a^{2}}{b^{2}}u^{2}(1+2u),
\end{equation}
which is the effective potential associated with Kerr solution, to
second order in $a$, as expected (see, for example, \cite{hartle}).

\begin{figure}[t!]
\begin{center}
\includegraphics[width=0.48\linewidth]{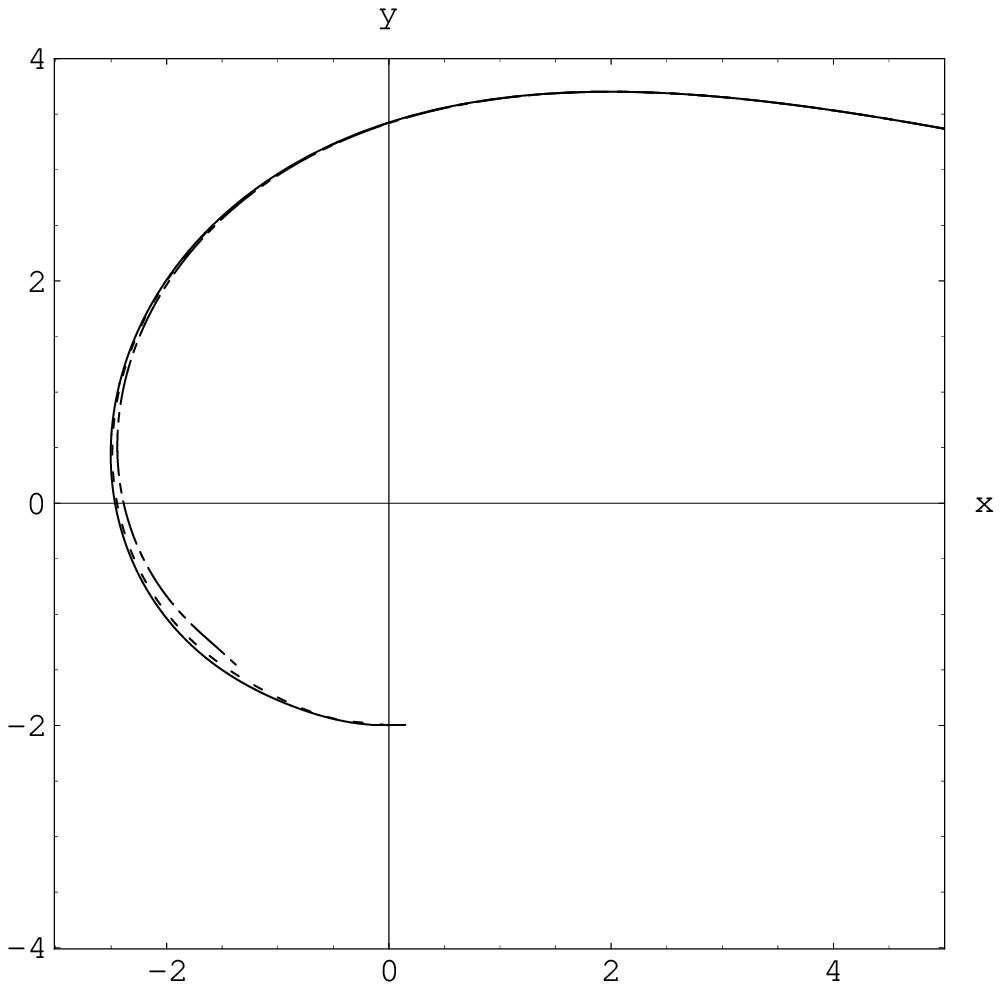} 
\hspace{0.5cm} 
\includegraphics[width=0.48\linewidth]{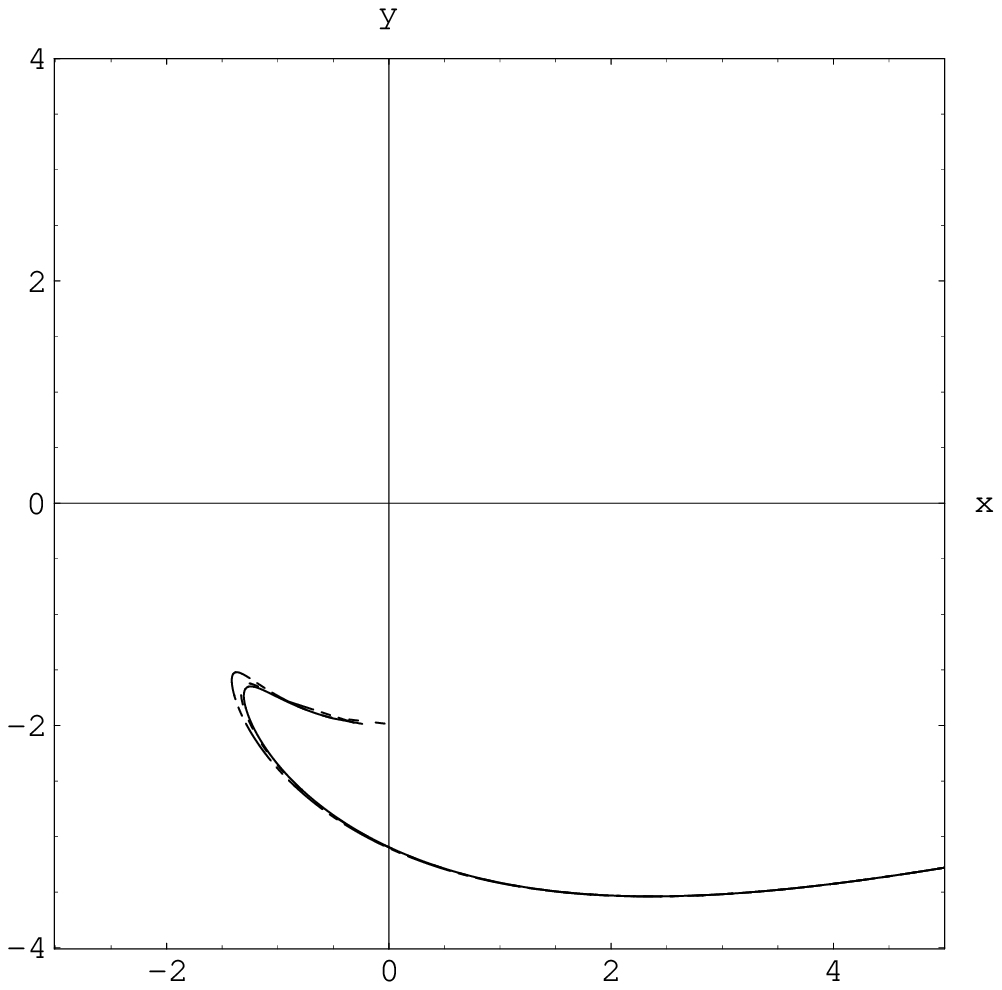}
\end{center}
\caption{Plot of equatorial photon orbits with impact parameters $b=4.5$
(left panel) and $b=-4.5$ (right panel), for a rotating black hole situated at the origin of coordinates with spin parameter $a=0.2$. The different curves correspond to CS parameters $\gamma=0.3$ (dashed-dotted line), $\gamma=0.15$ (dashed line) and $\gamma=0$ (solid
line). All quantities were adimensionalized with the mass of the
black hole.} 
\label{fig1}
\end{figure}

It is possible to obtain the equatorial orbits of photons around
black holes in GR or in CS modified gravity from the effective
potential $W_{\text{eff}}$. This potential has an unique extreme in the range $r>r_+$ (with $r_+$ the event horizon radius), which corresponds to a maximum. The behavior of the potential is similar to that in GR. The potential depends on the impact parameter $b$, so prograde and retrograde photons interact with different potentials. In fact, we have four types of possible equatorial orbits for photons:

\begin{itemize}
\item \emph{Scattering orbits}: photons that come from the infinity, reach the perihelion, and then scatter back to infinity. This kind of orbits happens when $1/b^{2}<W_{max}$, being $W_{max}$ the largest value of
the potential.

\item \emph{Falling orbits}: photons that come from the infinity, and then eventually fall into the black hole crossing the horizon. In this case, $1/b^{2}>W_{max}$.

\item \emph{Circular orbits}: unstable circular orbits with radius $r_{max}$, where $W(r_{max})=W_{max}$.

\item \emph{Falling orbits, initial position close to the horizon}: photons
that come from some initial radius $r_{0}$ such that $r_{+}<r_{0}<r_{max}$ , and end up falling into the black hole.
\end{itemize}
In this work, we concentrate on the first two types of orbits. The shape of the orbits can be obtained by numerical integration of $d\phi/dr$, which is the quotient between $d\phi/d\lambda$ and $dr/d\lambda$. The former derivative (to order $a\gamma ^2$) is given by
\begin{equation}
\frac{d\phi}{d\lambda}=bu^{2}+\frac{u^{3}a(2-bua)}{1-2u}-\frac{\pi u^{6}(70+120u+189u^{2})}{7(1-2u)}a\gamma^{2},
\end{equation}
while $dr/d\lambda$ is easily deduced from Eq. (\ref{dradtau}). In Figs. \ref{fig1} two examples are shown, which correspond to different trajectories of prograde and retrograde photons with $|b|=4.5$ around a black hole of $a=0.2$, for different values of $\gamma$. As we have previously mentioned, the main effect of the modified spinning solution of \cite{YP,konno} is producing a \emph{weaker} dragging of the inertial frames in the region close to the black hole, that manifests itself more clearly in the retrograde orbits of photons, which for growing values of the parameter $\gamma$ start to turn back \textit{later} dragged by the black hole.

In order to study the shadow cast by a spinning black hole, it is also
necessary to investigate the non-equatorial null geodesics. Let us move to
analyze this in the next subsection.

\subsection{General photon orbits}

To analyze the general orbits of photons around the black hole, we begin by
studying the separability of the Hamilton-Jacobi equation. Carter showed in
\cite{carter} that this separability is possible in the case of Kerr
geometry, using a third conserved quantity, often called Carter constant. In this section, we adopt the notation of \cite{chandra}.

The Hamilton-Jacobi equation, which determines the null geodesics for the
geometry given by the metric $g_{\mu\nu}$, is
\begin{equation}  \label{SHJ1}
\frac{\partial S}{\partial \lambda}=\frac{1}{2}g^{\mu\nu}\frac{\partial S}{\partial x^{\mu}}\frac{\partial S}{\partial x^{\nu}},
\end{equation}
where $S$ is the Jacobi action. The components of $g^{\mu\nu}$ are calculated here up to
order $a\gamma^2$. When the problem is \textit{separable}, the Jacobi action $S$ can be written in the form
\begin{equation}  \label{SHJ2}
S=\frac{1}{2}\delta \lambda - E t + L\phi + S_{r}(r)+S_{\theta}(\theta).
\end{equation}
The second term on the right hand side is related to the conservation of energy $E$, while the third term is related to the conservation of the angular momentum in the direction of the axis of symmetry $L$. In our case $\delta =0$ because we are dealing with null geodesics. Then, considering this \textit{ansatz} for $S$ in (\ref{SHJ1}), we get
\begin{equation}  \label{SHJ3}
2\frac{\partial S}{\partial \lambda}=0=g^{00}E^{2}-2g^{0\phi}E
L+g^{\phi\phi}L^{2}+g^{rr}\left(\frac{dS_{r}}{dr}\right)^{2} +
g^{\theta\theta}\left(\frac{dS_{\theta}}{d\theta}\right)^{2},
\end{equation}
where the right hand side of Eq. (\ref{SHJ3}) is calculated to order
 $a\gamma^2$. It might be instructive for the reader to compare this
expression with the one corresponding to the Kerr geometry,
calculated to second
order in $a$ (the exact expression for Kerr geometry can be found in \cite
{chandra}). Taking all this into account, the right hand side of Eq. (\ref{SHJ3}) can be expressed in the form
\begin{equation}  \label{CSmenosK}
0=F_{\text{SK}}-\Delta F_{\text{CS}},
\end{equation}
where $F_{\text{SK}}$ is the derivative of $S$ with respect to $\lambda$
for the case of the Kerr geometry, and to second order in $a$. $\Delta F_{\text{CS}}$ is a corrective term that appears from considering the rotating black hole solution with metric (\ref{metrica}). The expression for $\Delta F_{\text{CS}}$ to order $a\gamma^{2}$ is
\begin{equation}  \label{deltaCS}
\Delta F_{\text{CS}}=\frac{2 L E \pi u^{4} (70+120u+189u^{2})}{7(1-2u)}
a\gamma^{2}.
\end{equation}
It is easy to show that $F_{\text{SK}}$ is separable in two functions of $r$ and $\theta$, and the corrective term $\Delta F_{\text{CS}}$ is separable to order $a\gamma^{2}$, which is the order of the modified solution studied in the present work. Finally, it is not difficult to see that Eq. (\ref{CSmenosK}) is separable and gives two equations: one for $r$ and the other for $\theta$, whose expressions are, respectively,
\begin{equation}  \label{Sr}
\frac{1}{E^2}\left(\frac{dS_{r}}{dr}\right)^{2} =\frac{1}{\Delta }\left(-4 a \xi \frac{u}{1-2u}+\frac{1}{u^{2}(1-2u)}-\frac{u^{2}\left(4-\xi^{2}(1-2u)\right)}{(1-2u)^{2}}a^{2}+ \frac{2 \xi \pi u^{4} (70+120u+189u^{2})}{7(1-2u)}a\gamma^{2}-\eta-\xi^{2}\right)
\end{equation}
and
\begin{equation}  \label{Sth}
\frac{1}{E^2}\left(\frac{dS_{\theta}}{d\theta}\right)^{2}=a^{2}\cos^{2}(\theta)
-\xi^{2}\cot^{2}(\theta)+\eta,
\end{equation}
where $u=1/r$, $\xi=L/E$, $\eta=Q/E^{2}$, with $Q$ being the Carter constant, and $\Delta=u^{-2}-2u^{-1}+a^{2}$. In what follows it will be useful having defined the functions
\begin{equation}
R(r)={\Delta}^{2}\frac{1}{E^2}\left(\frac{dS_{r}}{dr}\right)^{2}
\end{equation}
and
\begin{equation}
\Theta(\theta)= \frac{1}{E^2}\left(\frac{dS_{\theta}}{d\theta}\right)^{2}.
\end{equation}
Then, by using Eq. (\ref{Sr}) the expression for $R(u)$, calculated
to  order $a\gamma^2$, is
\begin{equation}  \label{R}
R(u)=\frac{1-u^{2}(1-2u)(\xi^{2}+\eta)}{u^{4}}+\left(-4\frac{\xi}{u} +\frac{2 \xi \pi u^{2} (70+120u+189u^{2})}{7}\gamma^{2}\right)a+ \left(\frac{1+2u}{u^{2}}-\eta\right)a^{2}.
\end{equation}
Notice that replacing $\gamma =0$ in Eq. (\ref{R}) one recovers the expression for Kerr's metric (to second order in $a$). Regarding the function $\Theta(\theta)$, it is worth noticing that the equation (\ref{Sth}) is the same as in the Kerr geometry; therefore, it should satisfy the same conditions (see \cite{chandra} for details).

Finally, the Jacobi action $S$ reads
\begin{equation}  \label{Sfinal}
S=-Et+\xi \phi +\int_{r_0}^{r}\frac{\sqrt{R(r)}}{\Delta}dr+ \int_{\theta_0}^{\theta}
\sqrt{\Theta(\theta)}d\theta,
\end{equation}
where it was taken into account that both the energy $E$ and the angular momentum $L$ are conserved quantities (and consequently we have the conserved quantity $\xi=L/E$ as a parameter). Without losing generality, we can fix $E=1$.

The equations of motion corresponding to coordinates $r$ and $\theta$ can be simply obtained from $dS/dr=p_{r}=g_{rr}\dot{r}$ and $dS/d\theta =p_{\theta}=g_{\theta\theta}\dot{\theta}$. Then, combining these with (\ref{Sr}) and (\ref{Sth}), we have that
\begin{equation}  \label{rthpunto}
\frac{R(r)}{\Delta^{2}}=g_{rr}^{2}\dot{r}^{2}
\;\;
\mathrm{and}
\;\;
\Theta(\theta)=g_{\theta\theta}^{2}\dot{\theta}^{2}.
\end{equation}
The orbits with constant $r$ are those for which the conditions
\begin{equation}  \label{R0}
R(r)=0
\;\;
\mathrm{and}
\;\;
\frac{dR}{dr}(r)=0
\end{equation}
are satisfied. The values of the impact parameters $\xi$ and $\eta$ that
are compatible with these conditions determine the contour of the shadow of
the black hole. A detailed treatment of the shadow for (extremal) Kerr geometries can be found in \cite{chandra}, while other interesting related works are \cite{bambi,maeda}. In the case of rotating CS black holes, the
parameters $\xi$ and $\eta$ compatible with Eqs. (\ref{R0}) belong to two possible families, as in the case of Kerr geometry (see \cite{chandra}). However, one of these families is not consistent with the conditions that the function $\Theta(\theta)$ should satisfy. In our case, the family of allowed parameters is the one that in the limit $\gamma=0$ leads to the valid family for the Kerr geometry. Then, the expressions of $\xi$ and $\eta$ takes the form

\begin{eqnarray}  \label{xi}
\xi(u)&=&\xi_{\text{K}}(u)+\frac{\pi u^{2}}{7 a (1-2u)(u-1)} \left[
(1-3u)(140+90u+87u^{2}-945u^{3}) \frac{}{}\right.  \notag \\
&& + \left. \frac{2u^{2}(35+55u-101u^{2}-408u^{3}+102u^{4}+945u^{5})a^{2}}{1-2u} \right] \gamma^{2}
\end{eqnarray}
and
\begin{eqnarray}  \label{eta}
\eta(u)&=&\eta_{\text{K}}(u)-\frac{2\pi (1-3u)}{7 a^{2} (1-2u)(u-1)^{2}} 
\left[ (1-3u)(140+90u+87u^{2}-945u^{3}) \frac{}{} \right.  \notag \\
&& + \left.\frac{u^{2}(140+20u-233u^{2}-1143u^{3}+582u^{4}+1890u^{5})a^{2}}{1-2u} \right] \gamma^{2}
\end{eqnarray}
where
\begin{equation*}
\xi_{\text{K}}(u)=\frac{1-\Delta u-a^{2}u^{2}}{u(1-u)a},
\end{equation*}
\begin{equation*}
\eta_{\text{K}}(u)=\frac{4 \Delta u^{3}-(1-u)^{2}}{u^{4}(1-u)^{2}a^{2}}
\end{equation*}
are the expressions corresponding to the Kerr geometry.

\section{Black hole shadow}

As we have pointed out in the previous section, the allowed values for the
parameters $\xi$ and $\eta$ are those that determine the shadow of the black hole. If a black hole is situated between a source of light and an observer, the light reaches the observer after being deflected by the black hole gravitational field; but some part of the photons emitted by the source, those with small impact parameters, end up falling into the black hole, not reaching the observer. The apparent shape of a black hole is thus defined by the boundary of the shadow. To describe the shadow, we adopt the celestial coordinates:
\begin{equation}  \label{alpha}
\alpha=\lim_{r_{0}\rightarrow \infty}\left( -r_{0}^{2}\sin\theta_{0}\frac{d\phi}{dr}\right)
\end{equation}
and
\begin{equation}  \label{beta1}
\beta=\lim_{r_{0}\rightarrow \infty}r_{0}^{2}\frac{d\theta}{dr},
\end{equation}
where $r_{0}$ goes to infinity because we consider an observer very far from the black hole, and $\theta_{0}$ is the angular coordinate of the observer. The coordinate $\alpha$ is the apparent perpendicular distance of the image as seen from the axis of symmetry, and the coordinate $\beta$ is the apparent perpendicular distance of the image from its projection on the equatorial plane. If we calculate $d\phi/dr$ and $d\theta/dr$ from the metric given by Eq. (\ref{metrica}) and take the limit of a far away observer, we have that, as a function of the constants of motion, the celestial coordinates take the form
\begin{equation}  \label{alphapsi1}
\alpha=-\xi\csc\theta_{0}
\end{equation}
and
\begin{equation}  \label{beta2}
\beta=\frac{a^{2}+4(\xi^{2}+\eta)+a^{2}\cos2\theta_{0}-4\xi^{2}\csc^{2}
\theta_{0}} {4\sqrt{\eta-\xi^{2}\cot^{2}\theta_{0}}},
\end{equation}
where Eq. (\ref{rthpunto}) was used to calculate $u^{\theta}$.

\begin{figure}[t!]
\begin{center}
\includegraphics[width=0.48\linewidth]{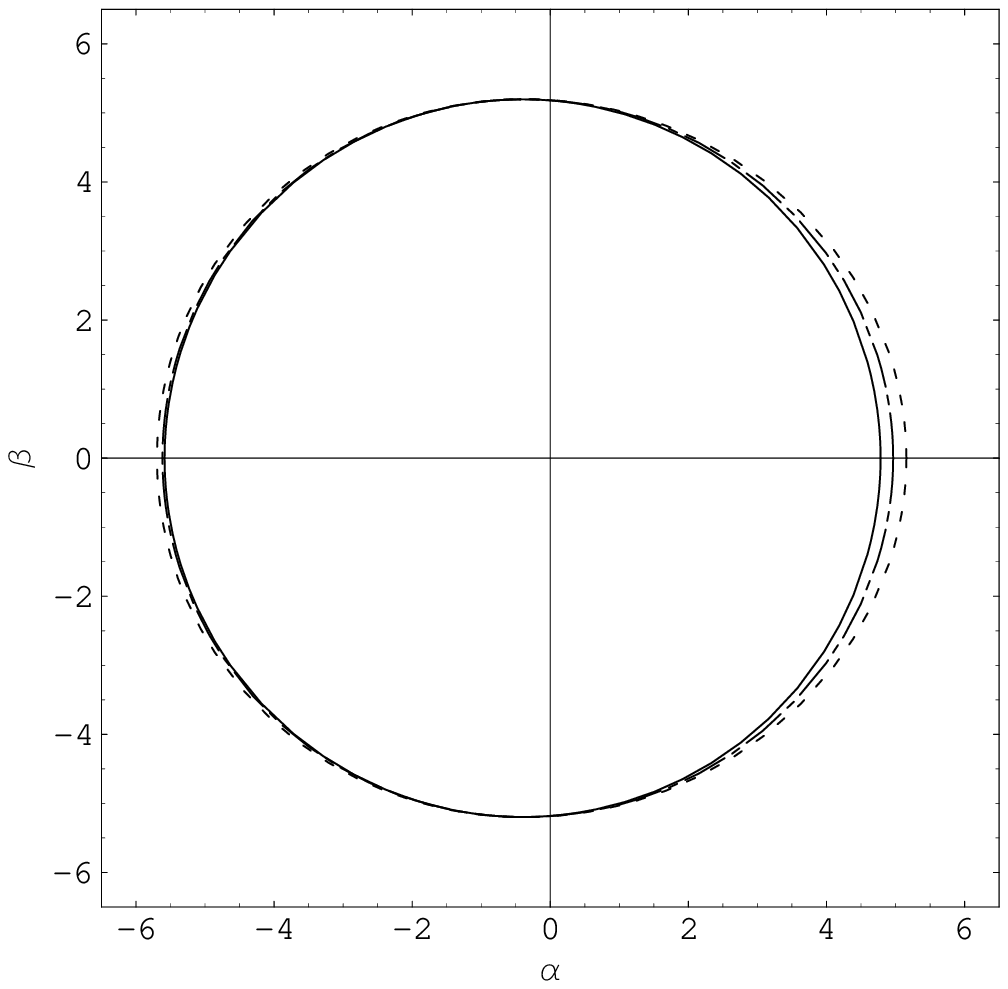} 
\hspace{0.5cm}
\includegraphics[width=0.48\linewidth]{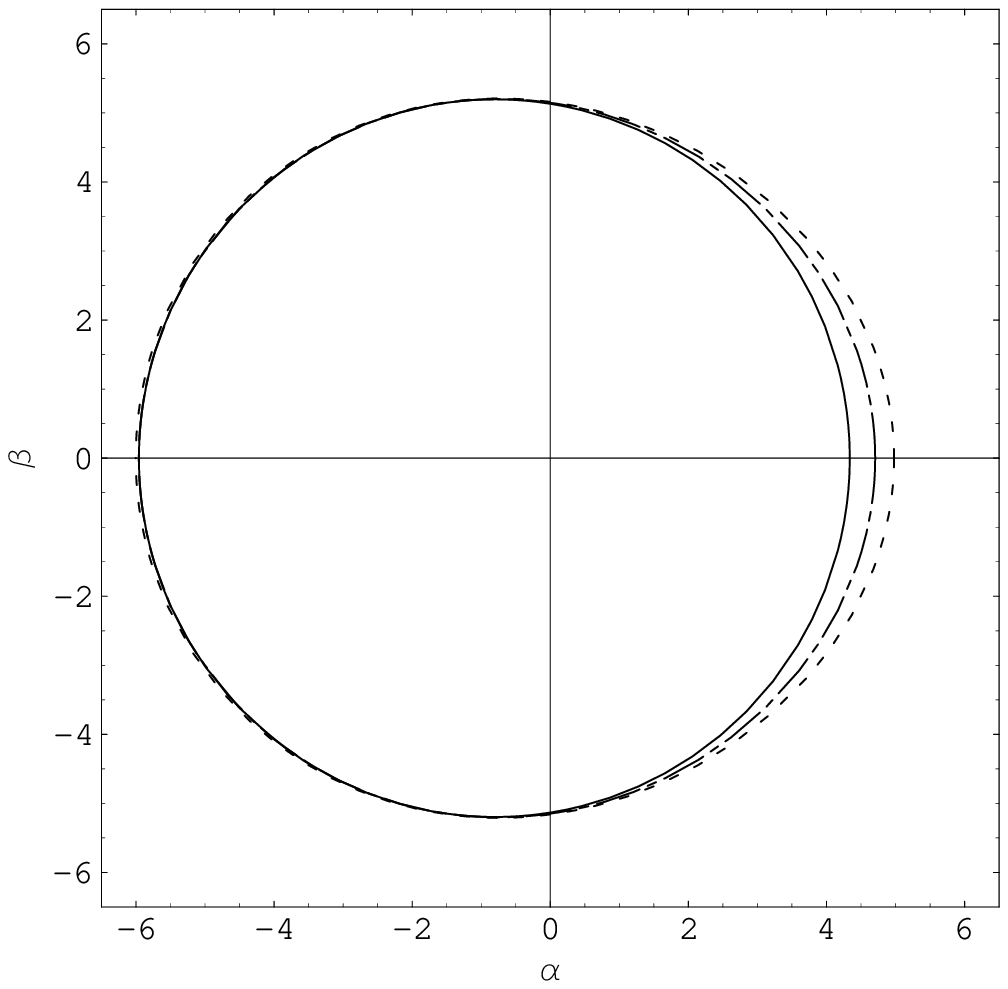}
\end{center}
\caption{Boundary of the shadow of a black hole situated at the origin of
coordinates with inclination angle $\theta _0=\pi /2$, and spin parameters $a=0.2$ (left panel) and $a=0.4$ (right panel). In both cases, the CS coupling parameters are $\gamma=0$ (Kerr case, solid curve), $\gamma=0.3$ (dashed-dotted curve) and $\gamma=0.4$ (dashed curve). All quantities were adimensionalized with the mass of the black hole (see text).}
\label{fig2}
\end{figure}

For the characterization of the form of the shadow, we adopt the observables defined in \cite{maeda}: the radius $R_{s}$ and the distortion parameter $\delta _{s}$. The quantity $R_s$ is the radius of a reference circle passing by three points: the top position $(\alpha_t, \beta_t)$ of the shadow, the bottom position $(\alpha _b,\beta _b)$ of the shadow, and the point corresponding to the unstable retrograde circular orbit when seen from an observer on the equatorial plane $(\alpha _r,0)$. The distortion parameter is defined by $D/R_s$, where $D$ is the difference between the
endpoints of the circle and of the shadow, both of them at the opposite side of the point $(\alpha _r,0)$, i.e. corresponding to the prograde circular orbit. The radius $R_s$ basically gives the approximate size of the shadow, while $\delta_s$ measures its deformation with respect to the reference circle (see \cite{maeda} for more details). If the inclination angle $\theta _{0}$ is independently known (see for example \cite{li-narayan}), precise enough measurements of $R_{s}$ and $\delta _{s}$ could serve, in principle, to obtain the rotation parameter $a$ and the CS parameter $\gamma$ (both adimensionalized with the black hole mass).

In the particular case where the observer is situated in such a way that the division line is in the equatorial plane of the black hole (for which the departures from GR are larger), the inclination angle is $\theta_{0}=\pi/2$
and we have simply
\begin{equation}  \label{alphapsi2}
\alpha=-\xi
\end{equation}
and
\begin{equation}  \label{beta3}
\beta=\sqrt{\eta}.
\end{equation}
These equations have implicitly the same form as for the Kerr's metric,  with the new $\xi $ and $\eta$ given by Eqs. (\ref{xi}) and (\ref{eta}) (a detailed calculation of the values of $\xi$ and $\eta$, and the expressions of the celestial coordinates $\alpha$ and $\beta$ as a function of the constants of motion for Kerr geometry, are given in \cite{vazquez}). For visualizing the \textit{shape} of the black hole shadow one needs to plot $\beta$ vs $\alpha $. In Fig. \ref{fig2} we show the contour of the shadows of black holes with rotation parameters $a=0.2$ and $a=0.4$ for some values of the CS coupling $\gamma $.

\begin{figure}[t!]
\begin{center}
\includegraphics[width=0.48\linewidth]{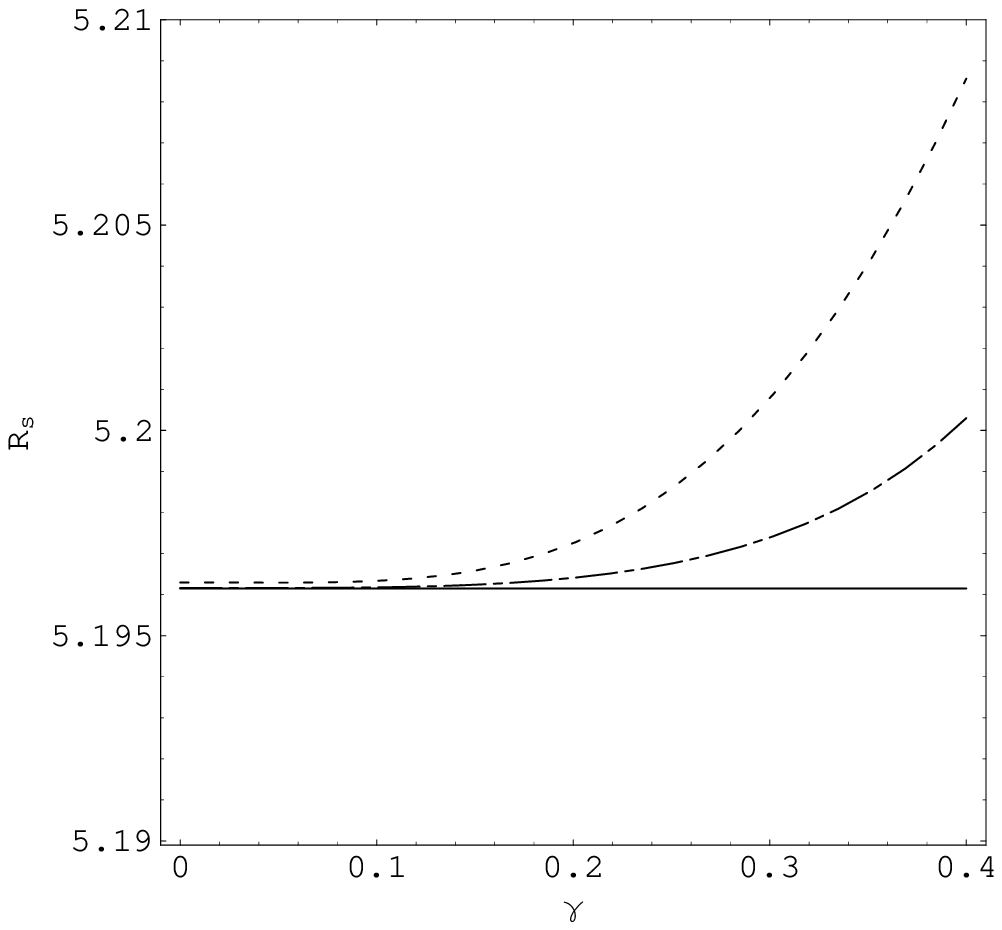} 
\hspace{0.5cm} 
\includegraphics[width=0.48\linewidth]{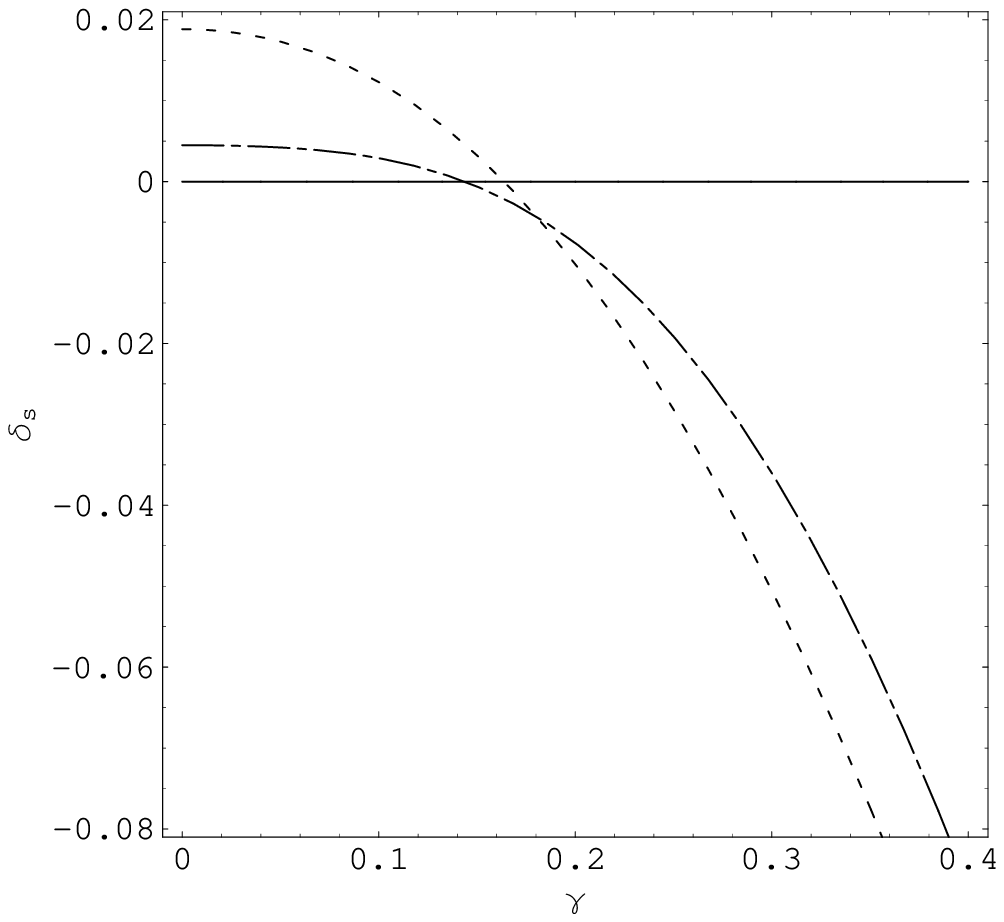}
\end{center}
\caption{Observables $R_{s}$ and $\delta _{s}$ as functions of the
CS coupling parameter $\gamma $, corresponding to the shadow of a
black hole situated at the origin of coordinates with inclination angle $\theta _0=\pi /2$, and spin parameters $a=0.2$ (dashed-dotted curves) and $a=0.4$ (dashed curves). For comparison, the values corresponding to the Schwarzschild black hole are $R_{s}=3\sqrt{3}\approx 5.19615$ and $\delta _{s}=0$ (solid curves). All quantities were adimensionalized with the mass of the black hole (see text).}
\label{fig3}
\end{figure}

The observable $R_s$ can be calculated from the equation
\begin{equation*}
R_{s}=\frac{(\alpha _t -\alpha_r)^2 + \beta_t ^2}{2|\alpha _t -\alpha_r|},
\end{equation*}
and the observable $\delta_s$ is given by
\begin{equation*}
\delta _s=\frac{\tilde{\alpha}_p - \alpha_p}{R_{s}},
\end{equation*}
where $(\tilde{\alpha}_p, 0)$ and $(\alpha_p, 0)$ are the points
where the reference circle and the contour of the shadow cut the horizontal
axis at the opposite side of $(\alpha_r, 0)$, respectively. In Fig. \ref{fig3} the observables $R_{s}$ and $\delta_{s}$ are shown as functions of $\gamma$. From Figs. \ref{fig2} and \ref{fig3}, we see that for a fixed value of $a$, the presence of the CS coupling $\gamma$ leads to a bigger shadow (larger $R_{s}$) than in the case of Kerr geometry, while a small value of $\gamma$ gives a less distorted shadow (smaller positive $\delta_{s}$) than for Kerr's; for large $\gamma$ the silhouette gets distorted in the opposite direction (negative $\delta_{s}$). For comparison, let us say that the non-rotating solution of CS gravity for any $\gamma$, i.e. the Schwarzshild black hole, has a circular shadow with radius $R_s=3\sqrt{3}\approx 5.19615$.

\section{Discussion}

In this work, we have studied the null geodesics corresponding to a slowly
rotating black hole in Chern Simons gravity, with a small coupling constant. We have shown that the photon orbits are separable as in the Kerr geometry. From the null geodesics we have found the shadow produced by the black hole. For a given inclination angle $\theta _0$, the deformation of the shape of the shadow with respect to a Schwarzschild black hole with the same mass would enable to extract information about the value of the angular momentum and the value of the CS coupling. This means that the aspect of the shadow allows to distinguish between the Kerr geometry and its CS modification. In this alternative theory, for a given rotation parameter $a$, the shadow is always larger, and less distorted than in GR when $\gamma$ is lower than a critical value, or distorted in the opposite direction if $\gamma $ exceeds that critical value. The key reason is that the effect of the CS term on the dragging is substantially stronger in the region close to the equatorial plane.

The values of $\gamma$ adopted in the plots were only for illustrative
purposes; the real values of $\gamma$ may be much smaller. The bound of the
CS coupling given in \cite{YP}, already mentioned in the Introduction, in the case of the adimensionalized parameter $\gamma$ can be rewritten in the form $\gamma < 1.4 \times 10^7 (M/M_{\odot})^{-2}$, where $M_{\odot}$ is the solar mass. For example, for a supermassive black hole with $M=10^6M_{\odot}$ we obtain $\gamma < 1.4 \times 10^{-5}$; on the other hand, for an intermediate mass one with $M=10^4 M_{\odot}$ we have $\gamma < 0.14 $, while for a stellar mass one with $M=10M_{\odot}$ the bound is $\gamma < 1.4 \times 10^{5}$. Then, in the case that the CS theory is a valid correction to GR, the known bound allows for a larger relative deviation from Kerr in the shadows of low mass black holes. This entails an extra observational difficulty, because the angles subtended by the shadows of stellar mass black holes --as seen from the Earth-- are much smaller than those corresponding to intermediate mass black holes or to the supermassive black hole Sgr A* at the Galactic center. The angular radius size of the shadow can be estimated from the Schwarzschild one with the same mass, which is given by $\theta _s = 3\sqrt{3}M/D_{o}$, with $D_{o}$ the distance from the observer to the black hole. It is easy to see that $\theta _s = 3 \sqrt{3}\times 10^{-5} (M/M_{\odot}) ( 1 \, \text{kpc}/D_{o})$ $\mu $arcsec. For Sgr A* we have $M=4.3 \times 10^{6}M_{\odot}$ and $D_{o}=8.3$ kpc \cite{guillessen} so we obtain $\theta _s=27$ $\mu $arcsec. For an intermediate mass black hole in a globular cluster, we can have $M\sim 10^{4}M_{\odot}$ and $D_{o}\sim 4$ kpc \cite{intmass}, then $\theta _s\sim 0.13$ $\mu $arcsec, while for a stellar size black hole we can take $M\sim 7 \,M_{\odot}$ and $D_{o}\sim 1.7$ kpc \cite{dunn}, giving $\theta _s\sim 2\times 10^{-4}$ $\mu $arcsec. Angular resolutions of the order of $1$ $\mu $arcsec are expected in the near future (see for example \cite{bozzareview}). The observation of the effect of the CS coupling on the shadow corresponding to the black hole in the vicinity of Sgr A* would be extremely difficult because of the very the small deformation allowed by the bound on $\gamma$. In the case of stellar size black holes, the main problem is not the bound on $\gamma $ but the small angular size of the shadow. It seems that the better candidates to observe the possible effects of the CS coupling on the shadows it would be the intermediate mass black holes in our galaxy if the existence of these objects is confirmed. In any case, the observation of the lensing effects due to a CS correction like those discussed in this paper will be inaccessible to current or near future technology.

The effects of the CS term on the shadows will be more prominent for large
values of the rotation parameter $a$ but, unfortunately, a solution of CS
gravity for every value of $a$ is not presently known.

\begin{acknowledgments}
This work was supported by ANPCyT, CONICET and UBA. We want to thank to Felix Mirabel, Leonardo Pelliza, and Gustavo Romero for suggesting useful references.
\end{acknowledgments}

\end{document}